\documentclass[%
twocolumn,%
showpacs,%
showkeys,%
preprintnumbers,%
amsmath,amssymb,%
page%
]{revtex4}
\usepackage{graphicx}
\usepackage{dcolumn}
\usepackage{bm}

\def\timeevol#1{\setbox1=\hbox{$\longmapsto$}\setbox2=\hbox{$
    \scriptstyle #1$}\copy1\kern-.5\wd1\kern-.5\wd2
    \raise-1.2\ht2\copy2\kern-.5\wd2\kern\wd1}

\begin{document}


\title{Altruistic Contents of Quantum Prisoner's Dilemma}
\author{
Taksu Cheon
}

\affiliation{
Laboratory of Physics, Kochi University of Technology,
Tosa Yamada, Kochi 782-8502, Japan
}

\date{November 8, 2004}

\begin{abstract}
We examine the classical contents of quantum games.
It is shown that a quantum strategy can be interpreted as
a classical strategy with effective density-dependent game matrices 
composed of transposed matrix elements.  
In particular,  successful quantum strategies in dilemma games 
are interpreted in terms of a symmetrized game matrix  
that corresponds to an altruistic game plan.
\end{abstract}

\pacs{02.50.Le, 03.67.-a, 87.23.Ge}
\keywords{game theory, quantum information, altruism}

\maketitle

%
%
Information processing with quantum mechanical bits, or qubits, 
has become a major area of research activities.
Among them, the quantum game \cite{ME99,EW99,OZ03} 
occupies a somewhat peculiar position.
This is because usual applications of game theory -- sociology, 
economics, evolutionary biology \cite{MA74, MS82, WE95} -- 
are so far removed from the realm of quantum mechanics.
It would be fair to say that quantum game theory has mostly been regarded as,
and indeed pursued as, an object of pure intellectual curiosity.
It is then somewhat puzzling to see quantum strategies ``succeed'' in some of the
long-standing  problems of classical game theory, whose solutions usually call for
rather involved concepts and techniques.
We cannot help wondering whether it is just a coincidence. 
Could it be that the quantum game theory amounts to an effective 
shorthand for advanced classical solutions?

In this article, we set out to give an answer to such questions
by examining in detail the working of quantum strategies in a two-by-two game.
When the quantum correlations are present, 
the classical separation of strategies of two players is in general not guaranteed.
Also, the correlations result in purely quantum component in the payoff functions.
However, it is pointed out that there is a particular choice of 
quantum strategies that ensures both classical separability
and intactness of classical payoff functions.
We show that this particular set of strategies, which we call pseudoclassical,
can be reinterpreted as a classical game played with a mixture of 
altruistic game plans \cite{AX84, BG98, CH03}.
We further show, through numerical examples on
the well-known case of the prisoner's dilemma,
that the altruism effectively incorporated in the pseudoclassical treatment
is indeed at the core of the success of quantum strategies. 

We consider a symmetric two strategy game,  described by 
a Hermitian operator $Q$ that is to be specified later, and
played by two players with quantum strategies $\left | \alpha \right >$ 
and $\left | \beta \right >$, both of which are linear combinations of 
basis strategy vectors $\left | 0 \right >$ and $\left | 1 \right >$.   We set
%
\begin{eqnarray}
\left | \alpha \beta \right> = {\cal U_{\alpha\beta}} \left | 0 0 \right>
\end{eqnarray}
The unitary rotation matrix ${\cal U_{\alpha \beta}}$ is given by
${\cal U_{\alpha \beta}}$ $= U_\alpha \bigotimes U_\beta$ where
$U_\alpha$ and $U_\beta$ act on the
qubits representing the first and the second players respectively. 
We adopt the notations
\begin{eqnarray}
U_\alpha = \begin{pmatrix}
      \alpha_0 &  \alpha_1   \\
     -\alpha_1^*  & \alpha_0^* \\
      \end{pmatrix}
,\quad
U_\beta = \begin{pmatrix}
      \beta_0 &  \beta_1   \\
     -\beta_1^*  & \beta_0^* \\
      \end{pmatrix}        ,
 \end{eqnarray}
with complex numbers satisfying the conditions
$|\alpha_0|^2+|\alpha_1|^2=1$ and $|\beta_0|^2+|\beta_1|^2=1$.
The payoff of the game to the first player is given by 
\begin{eqnarray}
\Pi(\alpha,\beta) =
\left < \alpha \beta  | Q  |  \alpha \beta  \right >  ,
\end{eqnarray}
where the quantum game operator $Q$ is set to be 
diagonal with respect to basis qubits of both players
%
\begin{eqnarray}
\left< i' j' | Q | i j \right> = \delta_{i',i} \delta_{j',j} A_{i j}
,
\end{eqnarray}
where  $A_{i j}$ is the  {\it classical} game matrix.
With the definitions $\alpha_i = \sqrt{x_i} e^{i\xi_i}$ 
and $\beta_i = \sqrt{y_i} e^{i\upsilon_i}$,
%
%
the payoff $\Pi$ depends only on the absolute values of $\alpha_i$ and $\beta_i$,
and takes the form
\begin{eqnarray}
\label{matp}
\Pi(x,y) = \sum_{i,j} x_i A_{i j} y_j
\end{eqnarray}
which is identical to the payoff of the purely classical 
mixed strategies that are described by the {\it strategy density vectors}
$x=(x_0, x_1)$ and $y=(y_0, y_1)$ with conditions $x_0+x_1=0$ and $y_0+y_1=1$.
Because of the symmetry of the system, 
the payoff for the player 2 is given by the conjugate payoff
\begin{eqnarray}
\label{cnjg}
\Pi^\dagger(x, y) = \Pi(y, x) .
\end{eqnarray}
Clearly, we have $\Pi(x,x)$ $=\Pi^\dagger(x,x)$, which simply means 
both players with same strategy earn the same payoff. 
%

A mixed Nash equilibrium $x^\star$ is defined by the condition
\begin{eqnarray}
\label{nash}
\left. \partial_{ x_i } \Pi(x, y) 
\right |_{x=y=x^\star} = 0 ,
\end{eqnarray}
with the implied assumption that it gives local maximum, not minimum.
This strategy $x^\star$ is the best response of a player against
an openent playing his/her best response. Since our game is symmetric, 
both players should play the same strategy $x^\star$ to obtain
the expected payoff  $\Pi(x^\star,x^\star)$.
There are, however, special cases where there is no solution to (\ref{nash})
within the valid range $0 \le x_i \le 1$, 
in which case, the best response becomes 
the {\it pure} Nash equilibrium, $x^\star=(0, 1)$ or $(1, 0)$, 
depending on the sign of $\partial_{x_i}\Pi(x,y)$.  
%

The Nash equilibrium, obtained as the result of individual pursuit of
optimality, is not always an ideal outcome for both players.  In fact, if we consider
$\Pi(x, x)$ as a function of strategy density $x$, and seek the value of $x$
that maximizes this function by
%
\begin{eqnarray}
\label{pareto}
\left. \partial_{ x_i } \Pi(x, x) 
\right |_{x=x^\circ}  = 0
,
\end{eqnarray}
the strategy vector $x^\circ$ is Pareto efficient, or both players are best off,
assuming this extremum is indeed a maximum. 
If a $x^\star$ is equal to $x^\circ$
the outcome of the game is described as Pareto efficient Nash equilibrium. 
Considering the relation
\begin{eqnarray}
\label{fullderiv}
\partial_{ x_i } \Pi(x, x) 
= \left . \partial_{ x_i } \{ \Pi(x, y) +\Pi^\dagger(x,y)\} \right |_{x=y}
,
\end{eqnarray}
we see that a game with the self-adjoint payoff has 
a Pareto efficient Nash equilibrium, namely,
\begin{eqnarray}
\label{selfcnjg}
x^\star = x^\circ
\quad {\rm if } \quad
 \Pi^\dagger(x,y) = \Pi(x,y) ,
\end{eqnarray}
because simultaneous constraints (\ref{nash}) and (\ref{pareto}) 
hold for this case.
When, on the other hand,
the Nash equilibrium $x^\star$ is remote from the Pareto efficient 
value $x^\circ$, the outcome of the game is less than optimal for both 
players.  This is the situation where the term {\it dilemma} is invoked, 
the representative of which is the well-known Prisoner's dilemma,
that has been the subject of much studies.
In fact, we might even say that
the search of the dynamics that brings Pareto efficient outcome to 
dilemma games constitute the bulk part of the recent works 
in evolutionary game theory.
Quantum strategies, which has been offered as an exotic 
alternative to the solution of dilemma games, 
deviate from the classical strategies with the introduction of 
quantum correlations.  Following the scheme of Eisert, Wilkens, 
and Lewenstein \cite{EW99}, we define
\begin{eqnarray}
{\cal J}_\gamma = e^{-i \frac{\gamma}{2} \sigma_2 \bigotimes\sigma_2}
\end{eqnarray}
to obtain a correlated state 
\begin{eqnarray}
{\cal J}_\gamma  \left | 0 0\right > 
= \cos{\frac{\gamma}{2}}  \left | 0 0 \right > 
+ i \sin{\frac{\gamma}{2}} \left | 1 1 \right > ,
\end{eqnarray}
and construct a correlated quantum strategy 
vector 
\begin{eqnarray}
\left | \Psi_{\alpha\beta}(\gamma) \right > 
= {\cal J}^\dagger_\gamma {\cal U}_{\alpha\beta} {\cal J}_\gamma
 \left | 0 0\right >  .
\end{eqnarray}
The payoff to the first player now becomes
\begin{eqnarray}
\Pi_\gamma(\alpha,\beta) = 
\left < \Psi_{\alpha\beta}(\gamma) | Q | 
          \Psi_{\alpha\beta}(\gamma) \right > .
\end{eqnarray}
With the split of $\cal U_{\alpha\beta}$ into real and imaginary components 
${\cal U}_{\alpha\beta}$ $= {\cal R}_{\alpha\beta} + i {\cal I}_{\alpha\beta}$,
we obtain
%
\begin{eqnarray}
\label{payg}
\Pi_\gamma(\alpha,\beta) 
= & &\!\!
\left < 0 0 \right | 
\left( 
{\cal R}^{\dagger}_{\alpha\beta} Q {\cal R}_{\alpha\beta}
+ {\cal I}^{\dagger}_{\alpha\beta} Q {\cal I}_{\alpha\beta} 
\right)
\left | 0 0 \right >
\\ \nonumber
- \sin^2{\gamma} & &\!\!\!\!\!\!
\left\{ {
  \left < 0 0 \right| {\cal I}^{\dagger}_{\alpha\beta} Q 
     {\cal I}_{\alpha\beta} \left | 0 0 \right >-
  \left < 1 1 \right | {\cal I}^{\dagger}_{\alpha\beta} Q 
    {\cal I}_{\alpha\beta} \left | 1 1 \right >  } \right\}
\\ \nonumber
-2\sin{\gamma}  & &\!\!\!\!\!\!
  \left < 1 1 \right | {\cal I}^{\dagger}_{\alpha\beta} Q 
    {\cal R}_{\alpha\beta} \left | 0 0 \right > .
\end{eqnarray}
We write the payoff for the first player in an analogous form to
the classical case, (\ref{matp}) using the strategy densities $x$ and $y$ of
the first and the second players;
\begin{eqnarray}
\Pi_\gamma(\alpha,\beta) =
\sum_{i,j} { x_i  B_{ij}(\gamma) y_j } .
\end{eqnarray}
The effective payoff matrix $B_{i j}(\gamma)$ is written as 
\begin{eqnarray}
B_{i j}(\gamma) = A_{i j} 
+ B^{exc}_{i j}(\gamma) + B^{cor}_{i j}(\gamma) ,
\end{eqnarray}
where the ``exchange'' contribution $B^{exc}_{i j}(\gamma)$ comming from
thesecond terms of (\ref{payg}) is given by
\begin{eqnarray}
B^{exc}_{i j}(\gamma) =
 - \sin^2\gamma\sin^2( \xi_i+\upsilon_j ) ( A_{i j} - A_{ \bar{i} \bar{j} } ) ,
\end{eqnarray}
and the ``correlation''  contribution $B^{cor}_{i j}(\gamma)$ coming from 
the last term is given by
\begin{eqnarray}
B^{cor}_{i j}(\gamma) 
\!\! &=& \!\!
(-)^{i+j} 2 \sin\gamma 
\sin( \xi_{\bar i}+\upsilon_{\bar j} ) \cos( \xi_{i}+\upsilon_{j} )
\\ \nonumber
& & \quad \times
\sqrt{ {x_{\bar i} y_{\bar j}} \over {x_i y_j} } 
A_{i j} .
\end{eqnarray}
Here the bars on top of the indices stand for the logical 
complementarity $\bar 0 = 1$ and $\bar 1 = 0$.
The correlation term $B^{cor}_{i j}$ has 
a singular strategy-density dependence.
In this form, it is obvious that
playing a given game specified by the matrix $\{ A_{i j} \}$ with quantum strategy is
formally equivalent to playing a related, but different game specified by 
the matrix $\{ B_{i j}(\gamma) \}$ with purely classical mixed strategy.
If we take the ``quantum strategy'' at its face value, both amplitudes and phases 
of $\alpha$ should be optimized to increase the payoff $\Pi_\gamma(\alpha,\beta)$.
Same apply for $\beta$ with $\Pi_\gamma(\beta,\alpha)$, and $\alpha$ and
$\beta$ would settle down at the common value corresponding to 
the full quantum Nash equilibrium.
However, such assumption would require a system consisting of
quantum agents making choice of strategies either with intelligence,
or under competitive evolutionary pressure.  We might argue that, at this point, such
approach is rather far fetched for real life ecosystems,
apart from artificial experimental realizations with quantum computational circuits.
In this article, with the purpose of analyzing the workings of quantum strategies,
we regard only the amplitudes $x_i$ and $y_i$ as the optimizing
variables and regard the angles $\xi_i$ $= \upsilon_i$ as external parameters, 
their equivalence being a natural reflection of the symmetry of players at
the final outcome.
Of all possible quantum strategies, there are four subsets under which 
contributions from $B^{cor}_{i j}$ disappears. 

First is the {\it trivial classical limit} $(\xi_0 = 0, \xi_1 = 0 )$ 
or $(\xi_0 = {\pi \over 2}, \xi_1 = {\pi \over 2} )$, at  which we have
$B_{i j}(\gamma) = A_{i j}$.

The second, more interesting case is what we call {\it pseudoclassical limit},
 $(\xi_0 = 0, \xi_1 = {\pi \over 2} )$
or $(\xi_0 = {\pi \over 2}, \xi_1 =0 )$, with which we have
\begin{eqnarray}
\label{pseudoa}
B_{i j}(\gamma) = \cos^2\gamma A_{i j} + \sin^2\gamma A_{j i} .
\end{eqnarray}
%
For this case,  the quantum payoff $\Pi_\gamma(x,y)$ calculated from (\ref{pseudoa})
is readily given by
\begin{eqnarray}
\label{pseudop}
\Pi_\gamma(x,y) = \cos^2\gamma \Pi(x,y) + \sin^2\gamma \Pi^\dagger(x,y) .
\end{eqnarray}
A simple classical interpretation of the pseodoclassical case exist 
in terms of  {\it altruism};  that is, a choice of $\gamma$ gives a specific 
{\it game plan}, or a style in the selection of strategies of a player, 
that incorporate the interest of other player.
For example, adopting $\Pi_{\pi \over 2}(x, y)$ $=\Pi^\dagger(x, y)$
 as one's payoff amounts to exchanging the role of two players, thus
identifying the payoff of the other party as one's own, or playing with the
totally altruistic game plan.  On the other hand,
 adopting $\Pi_{0}(x, y)$ $=\Pi(x, y)$ means 
just stick to one's own payoff as usual, or playing with selfish game plan.
Anything in between, ($0 < \gamma < \pi/2$), gives a game plan which
pursue varying mixture of selfish and altruistic payoff maximization. 

A notable consequence of (\ref{pseudop}) is  that, for a given common
strategy for both players, the payoff for quantum game is identical to that of 
classical game;  
%
\begin{eqnarray}
\label{equivg}
\Pi_\gamma(x,x) = \Pi(x,x)
.
\end{eqnarray}
The Nash equilibrium of quantum game play $x^\star_\gamma$ is now given by
\begin{eqnarray}
\label{nashgam}
\left. \partial_{x_i} \Pi_\gamma(x, y) 
\right |_{x=y=x^\star_\gamma} = 0
,
\end{eqnarray}
and both players end up obtaining 
the payoff 
$\Pi(x^\star_\gamma, x^\star_\gamma)$.

Among various game plans with different $\gamma $ values,
the one with $\gamma=\pi/2$ occupies a spacial place.  
Because of the equal mixture of ``selfish'' $\Pi(x,y)$ and 
``altruistic'' $\Pi^\dagger(x,y)$, we have self-adjointness for 
the payoff, $\Pi_{\pi\over 4}(x,y)$ $=\Pi^\dagger_{\pi\over 4}(x,y)$.  
Then, from (\ref{selfcnjg}) we obtain a Pareto efficient Nash equilibrium 
for $\gamma = \pi/4$, namely  

\begin{eqnarray}
x^\star_{\pi\over 4} = x^\circ ,
\end{eqnarray}
which is the primary result of this article.
Among the pseudoclassical game plans with a given $\gamma$, therefore,
$\gamma = \pi/4$ gives the optimal results for both players, and 
either $\gamma=0$ or $\gamma=\pi/2$ gives the less favorable results.
%

Thus, within psudoclassical limit, thanks to the
crucial identity (\ref{equivg}), we are able to interpret the 
the quantum strategies as an effective way to incorporate
the altruistic game plan, which can help improve the outcome of the
game toward the optimal result. 

There are two more cases for which 
the strategy density dependent term $B^{corr}_{i j}$ drops out.
%
One of them is the case of $(\xi_0 = {\pi \over 4},  \xi_1 = {\pi \over 4} )$ 
or $(\xi_0 = {3\pi \over 4},  \xi_1 ={3\pi \over 4} )$, with which we have
\begin{eqnarray}
\label{case3}
B_{i i}(\gamma) &=& \cos^2\gamma A_{i i} + \sin^2\gamma A_{\bar i \bar i} ,
\\ \nonumber
B_{i \bar i }(\gamma) &=& \cos^2\gamma A_{i \bar i} + \sin^2\gamma A_{\bar i i} .
\end{eqnarray}
Another case is $(\xi_0 = {\pi \over 4},  \xi_1 = {3\pi \over 4} )$ 
or $(\xi_0 = {\pi \over 4},  \xi_1 ={3\pi \over 4} )$, with which we have
\begin{eqnarray}
\label{case4}
B_{i i}(\gamma) &=& \cos^2\gamma A_{i i} + \sin^2\gamma A_{\bar i \bar i}, 
\\ \nonumber
B_{i \bar i}(\gamma) &=&  A_{i \bar i} .
\end{eqnarray}
Although these appear analogous to the altruistic pseudoclassical case,
there is no immediate interpretation in terms of the classical game, because of
the existence of exchanged components in diagonal matrix elements.  
Moreover, there is no such
relation as (\ref{equivg}) in neither of these cases, 
and we have {\it bona fide} quantum contribution to the payoff
$\Pi_\gamma(x,x)-\Pi(x,x)$, which, by its nature, is classically non-interpretable. 

As such, these cases can be thought of as belonging to the more general 
category of generic quantum cases 
that are given by arbitrary values for $(\xi_0, \xi_1)$.
We would then calculate Nash equilibrium
for a given set of values of $(\xi_0, \xi_1)$
from the full payoff $\Pi_\gamma(\alpha,\beta)$, in a fashion 
analogous to (\ref{nashgam}).
Other than possible numerical assessments, figuring out 
the meaning of quantum payoff for generic quantum cases is 
beyond the scope of the current work.  
In fact, it is unlikely that it is readily interpretable, until we have sound scheme
to place the game theory into the information theoretical framework,
which we are still lacking.
Conversely, the true role of the quantum game theory may be to lay foundation
for the information theoretic approaches of the game theory. 
%
%
%

\begin{figure}
\includegraphics[width=6.5cm]{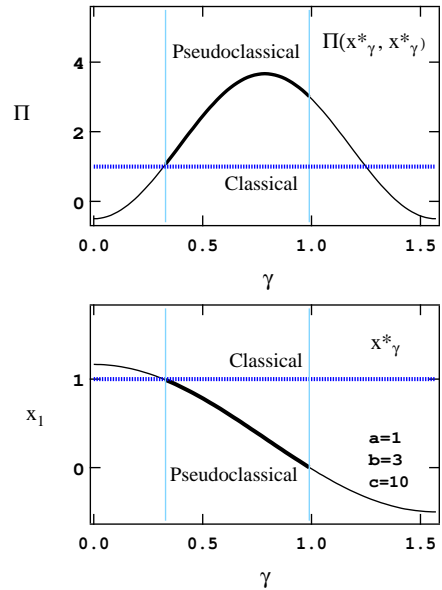}
\caption
{\label{fig1}
The Nash equilibrium strategy density ${x^\star_{\gamma}}_1$ (bottom) and the payoff
$\Pi(x^\star_\gamma,x^\star_\gamma)$  with the classical game (dotted line) and
pseudoclassical game (solid lines) for the game of Prisoner's dilemma (27). 
Horizontal axis is the parameter $\gamma$.
The classical game play leads to the least favorable
payoff $\Pi(x^\star)=a$ with ``defecting strategy''$x^\star_1 = 1$.  Among the 
pseudoclassical Nash equilibrium strategies $x^\star_\gamma$, $\gamma = \pi/4$
gives the Pareto efficient (best possible) payoff.}
\end{figure}
%
%
We illustrate our arguments with a numerical example on the Prisoner's dilemma.
With positive real numbers $a < b < c$, the classical game matrix is given by
%
\begin{eqnarray}
\{ A_{i j} \} 
= \begin{pmatrix}
     b &  0 \\
     c  & a \\
   \end{pmatrix} .
\end{eqnarray}
The Nash equibrium $x^\star=(1-x^\star_1, x^\star_1)$ 
and its payoff $\Pi(x^\star,x^\star)$ for the classical game is
%
\begin{eqnarray}
x_1^\star =1 
,\quad
\Pi(x^\star,x^\star) = a .
\end{eqnarray}
With the pseudoclassical quantum strategy density
$x_\gamma^\star=(1-x^\star_{\gamma 1}, x^\star_{\gamma 1}) $, 
they are given by 
\begin{eqnarray}
\label{pris1}
x^\star_{\gamma 1} 
\!\! &=& \!\!
 {1\over{a+b-c}} \left( b-c\sin^2 \gamma \right) ,
\\ \nonumber
\Pi(x^\star_\gamma,x^\star_\gamma) 
\!\! &=& \!\!
 {1\over{a+b-c}} \left( ab-{c^2\over 4}\sin^2 2\gamma \right)
,
\end{eqnarray}
for the case of $a+b<c$, and
\begin{eqnarray}
\label{pris2}
x^\star_{\gamma 1} = 0
,\quad
\Pi(x^\star_\gamma,x^\star_\gamma) = b ,
\end{eqnarray}
for the case of $a+b>c$.
An example of numerical results corresponding to 
the first case, (\ref{pris1}), is shown in FIG.1.

For separable, but fully quantum case (\ref{case3}), the corresponding 
results are obtained by replacing both $a$ and $b$ by their average $(a+b)/2$
in the formulae.   The results are
rather similar to the pseudoclassical case because of the existence of 
altruistic exchange components in (\ref{case3}).
For another separable, but fully quantum case (\ref{case4}),
there is no improvement over the classical game play
%

\begin{figure}
\includegraphics[width=6.5cm]{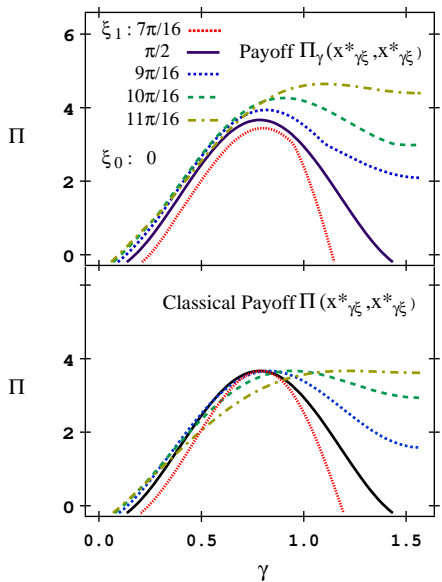}
\caption
{\label{fig2}
The classical payoff $\Pi(x^\star_{\gamma},x^\star_{\gamma})$ (bottom) and
full quantum payoff $\Pi_\gamma(x^\star_{\gamma},x^\star_{\gamma})$ (top)
are shown for the neighborhood of $(\xi_0, \xi_1) = (0, \pi/2)$.}
\end{figure}
%
In FIG.2, 
We show mumerical example of generic quantum cases of $\xi$ 
in the neighborhood of pseudoclassical case $\xi_0=0$, $\xi_1=\pi/2$.  
Both classical payoff $\Pi$ and full quantum payoff $\Pi_\gamma$ are
shown.
When we look at the classical payoffs, 
with generic ``quantum'' choice of strategies, $\gamma$-dependence
is changed from the pseudoclassical case. However, the essential 
ingredient of the successful strategy at high value of $\gamma$ --
mixture of altruism  -- is still intact.  
The story is similar with full quantum payoff functions.
Although the difference between the quantum payoff
and the classical payoffs are non negligible, 
the overall feature does not change very much.
While the results with only particular choices of angle parameters
are shown here, we note that these are rather representative 
ones whose characteristics are shared by
the results with other generic parameter values.  
We also add that in the fuller approach of ``complete quantum strategy'' implementation 
on the same Prisoner's dilemma including the optimization of angles \cite{EW99}, 
the stable quantum Nash equilibrium
is numerically found to coincide precisely with the pseudoclassical limit
in our terminology.

%
In conclusion, we have examined the source of the ``success'' of 
quantum strategies in dilemma games.  We have identified 
altruism, which is expressed in the symmetrization of the classical
game matrix, as the main cause.  In the process, 
the pseudoclassical limit of quantum strategies
with its intriguing characteristics  is uncovered.
\\

%
%
We are grateful to Professor K. Takayanagi, Professor T. Kawai,
and Prof. L. Hunter
for enlightening discussions.

%


\end{document}